\documentclass[prd,twocolumn,aps,showpacs,superscriptaddress,preprintnumbers,floatfix,amssymb]{revtex4}
\usepackage{epsfig}

\newcommand{\beqn}{\begin{eqnarray}}
\newcommand{\eeqn}{\end{eqnarray}}
\newcommand{\eq}[1]{(\ref{#1})}
\newcommand{\cL}{{\mathcal L}}
\newcommand{\bp}{{\mathbf p}}

\newcommand{\dd}{{\mathrm d}}

\newcommand{\Z}{{\mathbb Z}}
\newcommand{\res}{{\mathrm{res}}}

\newcommand{\logo}{\\ \vskip -23mm \leftline{\includegraphics[scale=0.3,clip=false]{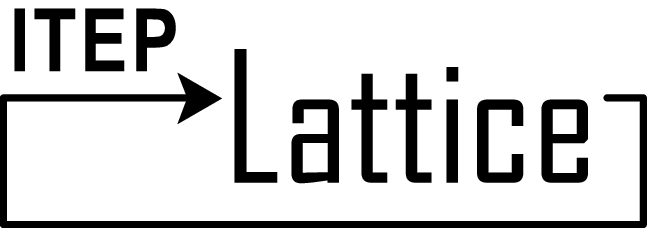}} \vskip 17mm}

\begin{document}

\title{Electric--magnetic asymmetry of the $A^2$ condensate \\
and the phases of Yang-Mills theory\logo}

\author{M.~ N.~Chernodub}\affiliation{ITEP, Bolshaya Cheremushkinskaya 25, Moscow, 117259, Russia}
\author{E.-M.~Ilgenfritz}\affiliation{Institut f\"ur Physik, Humboldt-Universit\"at
zu Berlin, Newton-Strasse 15, 12489 Berlin, Germany}

\preprint{ITEP-LAT/2008-11}
\preprint{HU-EP-08/16}

\begin{abstract}
We study the finite-temperature behavior of the $A^2$ condensate in the
Landau gauge of $SU(2)$ Yang--Mills theory on the lattice in a wide range
of temperatures. The asymmetry between the electric (temporal) and magnetic
(spatial) components of
this unconventional dimension--2 condensate is a convenient
ultraviolet-finite quantity which possesses, as we demonstrate, unexpected
properties. The low-temperature behavior of the condensate asymmetry suggests
that the mass of the lowest thermal excitation in the condensate is unexpectedly
low, about $200\,\mbox{MeV} \,$, which is much smaller than the glueball mass.
The asymmetry is peaking at the phase transition, becoming
    a monotonically decreasing function in the deconfinement phase. A symmetric point is reached in the
deconfinement phase at a temperature approximately equal twice the critical
temperature. The behavior of the electric-magnetic asymmetry of the condensate
separates the phase diagram of Yang-Mills theory into three regions. We suggest that
these regions are associated with the condensed, liquid and gaseous states
of the confining gluonic objects, the Abelian monopoles.
\end{abstract}


\date{May 23, 2008}

\maketitle

\section{Introduction}
\label{sec:introduction}

In a specific theoretical frame, non-perturbative features of QCD are reflected by
the existence of various non-vanishing local condensates.
The most famous vacuum condensates are the dimension--4 gluon condensate,
$\langle \alpha_s (G_{\mu\nu}^a)^2\rangle$,
and the dimension--3 quark condensate, $\langle \bar\psi \psi\rangle$.
The former characterizes the non-perturbative dynamics of strongly interacting
gluon fields while the latter -- in the formal limit of vanishing quark masses --
is an order parameter corresponding to the spontaneous breaking of
chiral symmetry. The local condensates enter the QCD sum rules as non-perturbative
power corrections having great significance for QCD
phenomenology~\cite{ref:sum:rules}.

The dimension--2 condensates represent somewhat unconventional vacuum condensates.
Indeed, it is impossible to construct
dimension--2 operator in QCD in a gauge-invariant and local manner, simultaneously.
One has to abandon either the condition of locality or the requirement of
gauge-invariance. Nonlocal but gauge invariant operators are useless from
the point of view of the operator product expansion: one cannot relate
a non-local operator to a Wilson coefficient corresponding to a power correction
using the dimensionality counting rule.
On the other hand, the local but gauge-dependent operators seem
to be useless because, as one
might think na\"ively, these operators cannot contribute to physical observables.
However, this conclusion seems to be not compelling as it was pointed out in
Ref.~\cite{ref:Gubarev:Zakharov:PRL,ref:Gubarev:Zakharov:PLB}.

The simplest dimension-2 operator in $SU(N_c)$ gauge theory
is~\cite{footnote1}:
\beqn
A^2(x) = \sum_{a=1}^{N_c^2 -1} \sum_{\mu=1}^4 A^a_\mu(x) A^a_\mu(x)\,.
\label{eq:primary}
\eeqn
This operator is not gauge invariant.
Therefore its expectation value can be understood in a two-fold way: one can
either average the operator over all possible gauge transformations or one
can evaluate the operator at a particular point of a gauge orbit.
The former choice leads to the unwanted non-locality while the latter is a
suitable option in particular gauges. The Landau gauge is defined as
the result of minimization
of the bulk averaged $A^2$--operator with respect to
gauge transformations. Thus, the extremum of the dimension--2 operator
gets a special meaning since $A^2(x)$ is a local operator in the Landau gauge.

The dimension--2 condensate enters the ultraviolet asymptotics of the gluon
propagator in the Landau gauge as a non-perturbative power-like $1/p^2$ correction.
Moreover, this condensate emerges also in QCD in the quark propagator and in various
vertices in the Landau gauge~\cite{ref:Pene:quark}. It is worthwhile
to remember that for a long time it was
assumed that due to asymptotic freedom the whole physics at short distances must
be described by exclusively perturbative physics. The appearance of the
dimension--2 operator in the ultraviolet regime is a signature
of mixing of non-perturbative and perturbative features of QCD: non-perturbative
effects are emerging at very short distances.

Besides the purely theoretical issue of mixing ultraviolet and infrared
physics, the appearance of the dimension--2 condensate plays a distinguished
r\^ole in QCD phenomenology because it is associated with non-standard power
corrections~\cite{ref:Gubarev:Zakharov:PLB,ref:Gubarev:Zakharov:PRL,ref:phenomenology}.
This motivates the wide interest in the subject,
such that the $A^2$-condensate was intensively studied both numerically and
analytically. Numerical simulations of lattice  $SU(3)$ Yang--Mills theory
at zero temperature indicate that the $A^2$ condensate is,
in fact, a large quantity~\cite{ref:Pene},
\beqn
\langle g^2 A^2 \rangle = {\left[1.64(15)\,\mbox{GeV} \right]}^2\,.
\label{eq:A2:num}
\eeqn

The energy scale of the dimension--2 condensate~\eq{eq:A2:num} is of the
order of the glueball mass~\cite{ref:Teper,ref:Bali}
\beqn
m_{O^{++}} = 3.52(12) \sqrt{\sigma(T=0)} = 1.55(5)\,\mbox{GeV}\,,
\label{eq:glueball}
\eeqn
if one takes the phenomenologically accepted value
$\sigma(T=0) = {[440\,{\mathrm{MeV}}]}^2$
for the string tension at zero temperature.
The coincidence does not look accidental.
Of course, the appearance of the dimension--2 condensate and the mass gap
generation in the non-Abelian gauge theory are both of non-perturbative nature.

Another non-perturbative phenomenon in non-Abelian gauge theory is the
confinement of color which is often linked with the mass gap generation
but the relevant scale is $\Lambda_{\mathrm{QCD}}
\sim
200\,\mbox{MeV}$.
A relation between the dimension--2 condensate and color confinement may also be
existent although the very reason is not clear at present.
A good playground to study the interrelations between the different mentioned
non-perturbative features would be a suitably chosen effective (toy) model.
The confinement of electric charges and the mass gap generation is understood,
for example, in the framework of an Abelian gauge model with a compact gauge
field, which can be considered as a certain limit of the $SO(3)$ Yang--Mills-Higgs model
(often called Georgi-Glashow model).
The compactness of the gauge field is related to presence of non-perturbative
objects, magnetic monopoles, the dynamics of which leads both to
confinement and to mass gap generation~\cite{ref:Polyakov}.

The four-dimensional version of the compact Abelian gauge model (sometimes
called ``compact electrodynamics'' or $cU(1)$ gauge theory) contains two
phases -- the confinement phase and the deconfinement phase -- separated by
a phase transition. At zero temperature this model is especially interesting
since the phases of the model are characterized by a single parameter,
the gauge coupling $g$. The confinement phase is located at strong coupling,
$g > g_c \sim 1$, while the deconfinement regime is associated with weak
coupling. The confinement phenomenon comes along with the mass gap generation
while in the deconfinement phase the mass gap shrinks to zero. Both mass gap
and charge confinement have the same, well understood~\cite{ref:Banks} roots
in the condensation of the Abelian monopoles (for a review see
Refs.~\cite{ref:Review,ref:review:vortices}).
The natural question to ask
is ``does an $A^2$-condensate appear in the compact Abelian model?''
If the answer is positive, what can we say about a possible
relation between confinement/mass gap generation and the emergence of the
$A^2$ condensate? These questions were addressed in
Ref.~\cite{ref:Gubarev:Zakharov:PRL}, where it was found that the non-perturbative
part of the $A^2$ condensate in the Landau gauge is a very good order parameter
for the confinement-deconfinement phase transition: this dimension--2 condensate
is non-vanishing in the confinement phase and vanishes in the deconfinement phase.
In the compact Abelian model the link between all three phenomena --
confinement, mass gap generation and emergence of the dimension--2
condensate is obvious: the primary reason of all these phenomena is monopole condensation.

Coming back to the Yang--Mills theory, one finds an essential difference
between this theory and the just discussed Abelian model: the transition
in the pure Yang--Mills theory is driven by thermal fluctuations and,
therefore, it happens at finite temperature. On the contrary, in the
mentioned simulations of the
compact Abelian gauge theory the transition is purely quantum, and thermal
fluctuations are not involved. In finite temperature Yang--Mills
theory (and in QCD as well), instead of having one dimension--2
condensate~\eq{eq:primary} one needs to define two types of it,
corresponding to time-like (``electric'') and to space-like (``magnetic'')
gauge bosons, separately. In the four-dimensional Euclidean space-time
corresponding to the imaginary-time formulation of the theory at $T \ne 0$,
one defines the electric and magnetic contributions, respectively:
\beqn
A^2_E & = & \frac{1}{N_c} {\mathrm{Tr}} \, A_4(x) A_4(x)\,,
\nonumber\\
A^2_M & = & \frac{1}{N_c} {\mathrm{Tr}} \, \sum_{i=1}^{3} A_i(x) A_i(x)\,,
\label{eq:secondary}
\eeqn
such that the full dimension--2 condensate at nonzero temperature is the sum of both,
\beqn
\langle g^2 A^2\rangle = \langle g^2 A^2_E\rangle + \langle g^2 A^2_M\rangle  \,.
\label{eq:sum:EM}
\eeqn
In Eq.~\eq{eq:secondary} we have not yet divided
the magnetic contribution by the number of spatial dimensions.
This would make $\langle g^2 A^2_E\rangle$ and $\langle g^2 A^2_M\rangle$
the dimension--2 condensate per one Lorentz component (or per three
Lorentz components) of the electric (magnetic) gluons in the Landau gauge.
We refine the conclusion of Ref.~\cite{ref:Furui:2006py} that the
$A^2$ condensate observed at zero
temperature is consistent with a vanishing condensate
in the deconfinement phase. In fact, the condensate at finite
temperature is characterized by the electric and magnetic components which,
as we show, are quite nontrivial.
To our knowledge, the eventual difference between the two
simplest dimension--2 condensates was not considered so far.
In this paper we fill this gap.

Before proceeding further one should comment about the ultraviolet
divergences of the dimension--2 condensate. In compact
electrodynamics~\cite{ref:Gubarev:Zakharov:PRL} at vanishing temperature
the condensate contains perturbative and non-perturbative parts, respectively,
\beqn
\langle g^2 A^2 \rangle = {\langle g^2 A^2 \rangle}_{\mathrm{pert}}
+ {\langle g^2 A^2 \rangle}_{\mathrm{NP}}\,.
\label{eq:decomposition}
\eeqn
The perturbative part is quadratically divergent whereas the non-perturbative
part is finite. The first one can be calculated trivially
in non-compact QED, the free theory of photons.
It is the non-perturbative part which serves as the order parameter
of the non-thermal phase transition in compact electrodynamics.

In QCD the situation is similar but not so
simple~\cite{ref:Pene,ref:Pene:testing,ref:Pene:artefacts}.
A linear decomposition~\eq{eq:decomposition} works as well up to
renormalization-related
logarithmic corrections. Then in Eq.~\eq{eq:decomposition}
${\langle
g^2 A^2 \rangle}_{\mathrm{pert}} \propto \Lambda_{\mathrm{UV}}^2$
and ${\langle g^2 A^2 \rangle}_{\mathrm{NP}} \propto \Lambda_{\mathrm{QCD}}^2$,
where $\Lambda_{\mathrm{QCD}}$ defines a typical QCD energy scale --
supplemented with a large prefactor according to Eq.~\eq{eq:A2:num} --
while $\Lambda_{\mathrm{UV}} \propto 1/a$ is an ultraviolet cutoff,
which in the case of lattice calculations is inversely proportional
to the lattice spacing $a$. The calculation of the non-perturbative part of
the condensate requires a renormalization which was implemented, for example,
in Ref.~\cite{ref:Pene}.

At finite temperature the decomposition~\eq{eq:decomposition} should
hold as well. It is well known that the finite-temperature corrections to
physical observables do not contain ultraviolet divergences.
However, the temperature corrections can be both of perturbative and
non-perturbative nature such that they can affect both the perturbative
and non-perturbative parts of the decomposition~\eq{eq:decomposition}.

In our numerical analysis we do not discriminate between perturbative and
non-perturbative parts.
Of course, we distinguish between corrections received by the spatial
(magnetic) and temporal (electric) gluons. On general grounds we write the
decomposition per Lorentz component in the form
\beqn
\langle g^2 A^2\rangle_E & = & \frac{1}{4} \langle g^2 A^2\rangle_0
                         + \langle g^2 A^2_E\rangle_T \, , \\
\langle g^2 A^2\rangle_M & = & \frac{3}{4} \langle g^2 A^2\rangle_0
                         + \langle g^2 A^2_M\rangle_T \, ,
\eeqn
where $\langle g^2 A^2\rangle_0$ is the dimension--2 condensate
at zero temperature while $\langle g^2 A^2_{E(M)}\rangle^T$
is the finite-temperature correction to the electric (magnetic)
dimension--2 condensate. These thermal corrections
are ultraviolet-finite at finite $T$ and
are vanishing at $T=0$. It is the zero-temperature condensate
$\langle g^2 A^2\rangle_0$ that contains a piece which is quadratically
divergent in the ultraviolet. In the limit of zero temperature magnetic
and electric components are equal and expressed naturally via the zero-temperature
condensate. In this limit the sum rule~\eq{eq:sum:EM} is also naturally restored.

We suggest to concentrate on the thermal corrections to the condensates,
$\langle g^2 A^2_E\rangle_T$ and
$\langle g^2 A^2_M\rangle_T$,
as probes of the thermal activity of the electric and magnetic components,
respectively, of the gluonic medium.
In this paper we compute the electric-magnetic asymmetry of the dimension--2
condensate
\beqn
\Delta_{A^2}(T) & = & \langle g^2 A^2_E\rangle - \frac{1}{3} \langle g^2 A^2_M\rangle
\nonumber\\
& \equiv & \langle g^2 A^2_E\rangle_T - \frac{1}{3} \langle g^2 A^2_M\rangle_T
\nonumber\\
& = &
{\langle g^2 A^2_4 \rangle}_T
- \frac{1}{3}\sum_{i=1}^3 {\langle g^2 A^2_i \rangle}_T \,.
\label{eq:CA:definition}
\eeqn
The asymmetry~\eq{eq:CA:definition} plays a special role
since the quadratically divergent
zero-temperature components of both electric and magnetic condensates
cancel out. Thus the asymmetry~\eq{eq:CA:definition} is
a finite-valued quantity in the ultraviolet regime.

We have calculated the asymmetry~\eq{eq:CA:definition} using numerical
simulations of Yang--Mills lattice gauge theory in the Landau gauge.
Obviously, at zero temperature (symmetric lattice)
all Lorentz components of the gauge field $A_\mu$ contribute equally to the $A^2$
condensate because of the approximate $O(4)$ rotational symmetry
(actually $H(4)$ hypercubic symmetry) satisfied by
the Euclidean lattice. At finite temperature this rotational symmetry is broken
down to (approximate) $O(3)$ spatial symmetry.
Space and time coordinates are no more equivalent since,
in the imaginary time formalism used to study the field
in a thermodynamic equilibrium state,
a finite temperature~$T$ is implemented via
compactification of the (imaginary) time direction to a circle of length
$L_t=1/T$, while the other (spatial) directions $L_s$ are still infinite,
or at least $L_s \gg L_t$. Thus, the fluctuations of the temporal (electric)
$A_4$ and spatial (magnetic) $A_i$ ($i=1,2,3$) components of the gluon fields
must be different in general.

The structure of this article is as follows.
In the following Sect.~\ref{sec:theoretical_expectations} we describe the
theoretical expectations.
In Sect.~\ref{sec:numerical_simulations} we
report our lattice simulations for $SU(2)$ pure gauge theory.
Sect.~\ref{sec:monopoles} contains a discussion relating our findings to
monopoles and confinement.
We draw conclusions and define further routes in Sect.~\ref{sec:conclusions}.

\section{Theoretical expectations}
\label{sec:theoretical_expectations}

The low- and high-temperature asymptotics of the electric-magnetic asymmetry of the
dimension--2 condensate $\Delta_{A^2}$ can be guessed from general arguments:
\beqn
\Delta_{A^2}(T) \propto
\left\{
\begin{array}{lcl}
e^{ - \frac{m_{\mathrm{gl}}}{T} } & \quad & T \ll T_c\,, \\
T^2 & & T \gg T_c\,.
\end{array}
\right.
\label{eq:asymptotics}
\eeqn
It is tempting to equate the mass parameter $m_{\mathrm{gl}}$ with the mass of the lowest glueball:
\beqn
m_{\mathrm{gl}} = m_{O^{++}}\,, \qquad\qquad \mbox{[na\"ive expectation]}\,.
\label{eq:mass:naive}
\eeqn
In Eq.~\eq{eq:asymptotics} a polynomial prefactor
at low temperatures and possible logarithmic corrections at high
temperatures are omitted.
The high temperature asymptotics in Eq.~\eq{eq:asymptotics}
is determined for dimensional arguments.
In the result of lattice simulations we will
show that the dimensional argument is correct, as expected.

Common wisdom says that the low-temperature asymptotics of any
thermodynamic system is determined by properties of the lowest excitation.
The lowest excitation in the $SU(N_c)$ gauge theory is a glueball,
which has a non-zero mass due to the phenomenon of the mass gap generation.
According to Ref.~\cite{ref:Teper}, in the $SU(3)$ gauge theory
$$m_{O^{++}} = 3.52(12) \sqrt{\sigma(T=0)}\,.$$
According to the Bielefeld group, Ref.~\cite{ref:Karsch:PRL,ref:Karsch:NPB}
the finite temperature phase transition in pure $SU(3)$ gluodynamics happens
at $$T_c = 0.629(3) \sqrt{\sigma(T=0)}$$ such that $$m_{O^{++}} \approx 5.6 \, T_c\,.$$
Given that $m_{O^{++}} \gg T_c$ the low-temperature asymptotics should
hold true not only at $T \ll T_c$, but also at temperatures close to the phase
transition temperature. As we will see below, in Yang-Mills theory
the exponential form of the low-temperature
asymptotics is correct while, unexpectedly, the ``natural'' identification of mass
scale~\eq{eq:mass:naive} turns out to be wrong.
Moreover, the low-temperature asymptotics~\eq{eq:asymptotics}
of the asymmetry turns out to be also incorrect in certain simple models
indicating that the exponential low-temperature asymptotics~\eq{eq:asymptotics}
is not valid in a general case.

We begin the discussion of the asymmetry in the simplest case of free
photodynamics and then for the Abelian Higgs model. After that we
turn to the case of Yang--Mills theory.

\subsection{Example of photodynamics}
\label{sec:photodynamics}

Photodynamics is the theory of the free Abelian gauge field.
Since the theory does not
contain any dimensional scale and is not able to generate
a scale by itself like QCD, the electric--magnetic asymmetry of the
$A^2$ condensate -- which is the dimension-2 quantity $\Delta_{A^2}(T)$ --
must be proportional to $T^2$ for dimensional reasons.

Since the Lagrangian of photodynamics is quadratic,
\beqn
\cL_{\mathrm{phot}} = \frac{1}{4} F^2_{\mu\nu}\,,
\qquad
F_{\mu\nu} = \partial_\mu A_\nu -  \partial_\nu A_\mu\,,
\label{eq:photodynamics}
\eeqn
it is easy to calculate the photon correlation function (propagator)
in momentum space
\beqn
\langle \tilde{A}_\mu(p) \tilde{A}_\nu(-p) \rangle = D_{\mu\nu}(p)\,,
\eeqn
where $p = (\bp, p_4)$ is the 4-momentum, while the relation between
the photon fields in coordinate and momentum spaces is given by Fourier
transformation. At zero temperature the relation is
\beqn
A_\mu(x) = \int \frac{\dd^4 p}{{(2\pi)}^4} e^{i (p, x)} \tilde{A}_\mu(p)\,,
\eeqn
while at finite temperature it is given by the formula
\beqn
A_\mu(x) = \int \frac{\dd^3 p}{{(2\pi)}^3} T \sum_n
e^{i (\bp, {\vec x}) + i \omega_n x_4} \tilde{A}_\mu(\bp, \omega_n) \,,
\eeqn
where
\beqn
\omega_n = 2 \pi n T
\label{eq:Matsubara}
\eeqn
are the Matsubara frequencies.

Due to absence of interactions the form of (zero temperature) photon propagator
\beqn
D_{\mu\nu}(p) = D(p^2) \Bigl( \delta_{\mu\nu} - \frac{p_\mu p_\nu}{p^2} \Bigr)\,,
\label{eq:photon:propagator}
\eeqn
is also valid at finite temperature. The propagator~\eq{eq:photon:propagator}
is parameterized by a single propagator function (here $p^2 = \bp^2 + p_4^2$):
\beqn
D(p^2) \equiv D^{\mathrm{photo}}(p^2) = \frac{1}{p^2}\,.
\eeqn

The asymmetry of the dimension-2 condensate in
photodynamics is calculated in Appendix~\ref{sec:appendix:photo}:
\beqn
\Delta^{\mathrm{free}}_{A^2}(T) & = & \frac{1}{3 \pi^2} \int\limits_0^\infty \dd p \, p^2 f_T(p)
\Bigl\{\frac{1}{2 p} - \frac{1}{T} \Bigl[1 + f_T(p) \Bigr] \Bigr\} \nonumber\\
& = & - \frac{T^2}{12} \, .
\label{eq:CA:photon}
\eeqn
As the definition of the asymmetry we took Eq.~\eq{eq:CA:definition} with one-component gauge field, $N_g=1$.
We have also omitted the overall coefficient, electric charge squared, $g^2=e^2$, in the definition~\eq{eq:CA:definition}
in order to simplify the expressions below. This coefficient can easily be restored: it should enter
all the analytic results of this Section for the asymmetry as just the proportionality coefficient.
The proportionality of the asymmetry to the squared temperature in Eq.~\eq{eq:CA:photon}
is quite
obvious due to dimensionality reasons. The fact that the asymmetry~\eq{eq:CA:photon} is negative tells
us that in the absence of the interactions the fluctuations of the magnetic (spatial) photons are dominating
the fluctuations of the electric (temporal) photons.

\subsection{Example of Abelian Higgs model}
\label{sec:AHM}

Now we turn to a more complicated case, adding a charged scalar field to photodynamics.
This system is described by the Abelian Higgs model with the Lagrangian
\beqn
\cL_{\mathrm{AHM}} = \frac{1}{4} F^2_{\mu\nu} + |(\partial_\mu + i e A_\mu) \phi|^2 + \lambda (|\phi|^2 - \eta^2)^2\,,
\label{eq:AHM}
\eeqn
and controlled by the gauge coupling $e$ and the quartic coupling $\lambda$. The overall dimensional scale
is fixed by the position $\eta$ of the minimum of the potential for the Higgs field $\phi$.

In the Higgs phase the photon has a mass $m = e \eta$ due to spontaneous symmetry breaking caused
by the condensation of the scalar Higgs field. At zero temperature the photon propagator
is described by Eq.~\eq{eq:photon:propagator} with the
single propagator function
\beqn
D(p) = D^{\mathrm{AHM}}(p) = \frac{1}{p^2 + m^2} \,.
\label{eq:Dp:AHM}
\eeqn

In the case of the London limit, $\lambda \to \infty$,
with the Higgs field being infinitely massive, the latter
is not thermally excited if the system is subject to finite temperatures.
As the result, the Higgs loops do not contribute to the photon polarization tensor.
The difference between
spatially longitudinal and spatially transverse photons is absent in this case
as they share the same propagator function~\eq{eq:Dp:AHM}.
Then the Lorentz structure at finite temperature of the gauge propagator
stays the same as in Eq.~\eq{eq:photon:propagator}.

The asymmetry for the London limit is calculated in Appendix~\ref{sec:appendix:AHM}:
\beqn
\Delta^{\mathrm{AHM}}_{A^2}(T,m) & = & \frac{4}{3 m^2} \Bigl[\varepsilon(T,m) - \varepsilon(T,m=0)\Bigr]
\nonumber\\
& & - \frac{1}{3} \Sigma(T,m) \, ,
\label{eq:CA:AHM}
\eeqn
where
\beqn
\varepsilon(T,m) & = & \int \frac{\dd^3 p}{(2 \pi)^3} \sqrt{\bp^2 + m^2}~f_T(\sqrt{\bp^2 + m^2})
\label{eq:energy}
\eeqn
and
\beqn
\Sigma(T,m) & = & \int \frac{\dd^3 p}{(2 \pi)^3} \frac{1}{\sqrt{\bp^2 + m^2}}~f_T(\sqrt{\bp^2 + m^2})\,.
\quad
\label{eq:sigma}
\eeqn

\begin{figure}[!htb]
\begin{center}
\includegraphics[width=8cm]{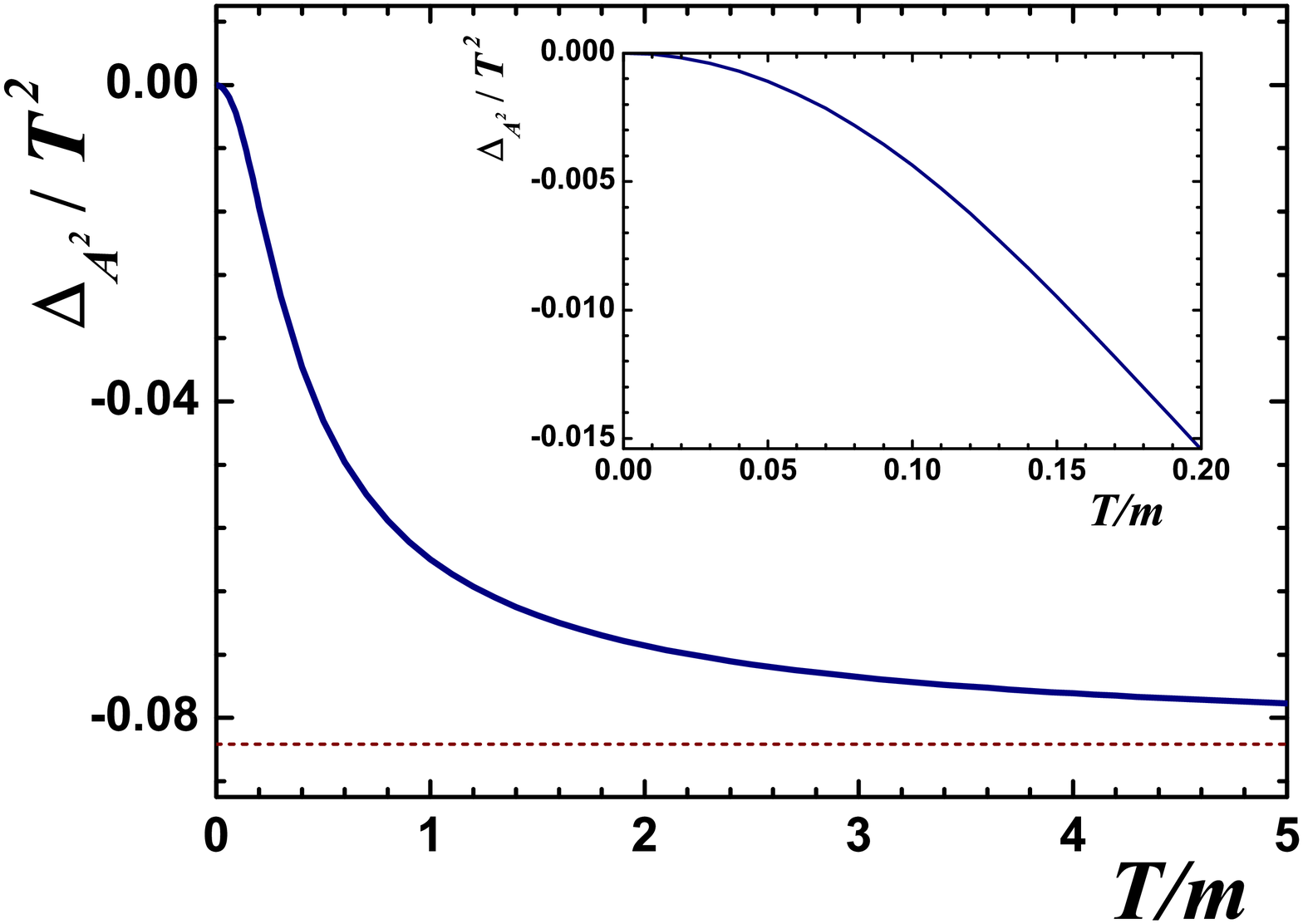}
\end{center}
\caption{(Color online) The normalized electric-magnetic asymmetry of the $A^2$ condensate,
$\Delta_{A^2}(T,m)/T^2$, as a function of the normalized temperature, $T/m$,
in the Abelian Higgs model. The horizonal dashed line in the plot shows the
high-temperature limit~\eq{eq:AHM:lim:high}, for $T \gg m$, which recovers
the case of photodynamics. The inset illustrates the low-temperature behavior
of the condensate asymmetry~\eq{eq:AHM:lim:low}.}
\label{fig:CA:AHM}
\end{figure}

In Figure~\ref{fig:CA:AHM} we show the asymmetry~\eq{eq:CA:AHM} divided by the
temperature squared, $T^2$, as a function of the temperature~$T$ expressed
in units of the gauge boson mass~$m$.

In the massless limit, $m \to 0$, the asymmetry~\eq{eq:CA:AHM} can be
calculated exactly:
\beqn
\lim_{m \to 0} \Delta^{\mathrm{AHM}}_{A^2}(T,m) = - \frac{T^2}{12} \,.
\label{eq:limit}
\eeqn
The comparison of Eq.~\eq{eq:limit} with Eq.~\eq{eq:CA:photon} shows
that the $A^2$--asymmetry is an analytic function of the photon mass at $m = 0$:
\beqn
\lim_{m \to 0} \Delta^{\mathrm{AHM}}_{A^2}(T,m) = \Delta^{\mathrm{AHM}}_{A^2}(T,0) \equiv \Delta^{\mathrm{photo}}_{A^2}(T) \, .
\eeqn
This result is, in fact, not guaranteed from the beginning
-- as we describe this in the Appendices --
since the exactly massless case involves the calculation of a residue at
the double pole while the massive theory has always two isolated single
poles in the asymmetry.

Equation~\eq{eq:limit} corresponds, in fact, to the leading term in the asymmetry
in the high temperature limit, $T \gg m$. Supplementing this result with the
low-temperature
expansion we get the leading terms in high- and low-temperature limits, respectively:
\beqn
\Delta_{A^2}(T,m) & = & - \frac{T^2}{12}  + \dots  \qquad\qquad (T \gg m)\,,
\label{eq:AHM:lim:high}
\\
\Delta_{A^2}(T,m) & = & - \frac{2 \pi^2}{45} \frac{T^4}{m^2} + \dots  \qquad (T \ll m)\,,
\label{eq:AHM:lim:low}
\eeqn
where the ellipsis denote subleading contributions.

As we have expected the difference between the
electric and magnetic components of the condensate in the Abelian Higgs model (AHM) is
vanishing in the zero temperature
limit according to Eq.~\eq{eq:AHM:lim:low}.
What comes unexpected is that
at low temperatures the asymmetry is suppressed
polynomially (by the fourth power of the temperature, $T^4$)
and not exponentially as one
would have expected from general arguments presented
in the beginning of this Section.
Indeed, the Abelian Higgs model is a theory with a finite mass gap,
and therefore one could expect
that thermodynamical contributions to any quantity would
be suppressed exponentially in the low
temperature limit by a factor $\exp\{ - m/T \} $.
Our analytical calculation shows this is not the case.

Technically, the unusual polynomial behavior $\Delta_{A^2}(T,m) \propto T^4$
of the asymmetry
at low temperatures is due to the massless, $m=0$, term in Eq.~\eq{eq:CA:AHM}.
This term corresponds to a massless longitudinal degree of freedom
while all other terms are massive
such that they are suppressed exponentially as $T \ll m$.
The appearance of this term can be traced in the integrand~\eq{eq:b2}
of the integral representation of the asymmetry~\eq{eq:inter:a2} as this integrand
contains a massless pole.
The massless pole, in turn, appears due to the fact that the propagator of the gauge boson~\eq{eq:photon:propagator}
contains a $1/p^2$ term which does not give any contribution
in explicitly transverse gauge-invariant expressions such as, for
example, the correlator of two field strength tensors.
However, the $A^2$-propagator {\it does} incorporate an infrared $1/p^2$ term, which
gives rise to a polynomial behavior unless the propagator function $D(p)$
is vanishing in the infrared.  Both in
photodynamics and in the Abelian Higgs model the
photon propagator is either divergent or finite in the infrared limit such that
the zero temperature limit of the asymmetry is not exponentially suppressed
contrary to our na\"ive expectation~\eq{eq:asymptotics}.

\subsection{Distinguishing longitudinal and transverse photons}
\label{sec:preparing_for_YM}

Contrary to the considered examples,  at finite temperature the
gluon propagator in the Landau gauge is in general parameterized by two
propagator functions. These are the transverse (or, ``magnetic'')
propagator $D_T$ and the longitudinal (or, ``electric'') propagator
$D_L$. In momentum space,
\beqn
D^{ab}_{\mu\nu}(\bp,p_4) =
\delta^{ab} \Bigl[P^T_{\mu\nu}\, D_{T}(\bp,p_4)
                + P^L_{\mu\nu}\, D_{L}(\bp,p_4) \Bigr] \, .
\label{eq:propagator:Tneq0}
\eeqn
Here $P^T_{\mu\nu}$ and $P^L_{\mu\nu}$ are,
respectively, projectors onto spatially transverse and spatially longitudinal
directions~\cite{ref:Kapusta}, with
\beqn
P^T_{\mu\nu} = (1-\delta_{\mu4})~(1-\delta_{\nu4})\,
\left(1 - \frac{p_\mu \, p_{\nu}}{\bp^2} \right)\,,
\eeqn
satisfying the relation
\beqn
P^T_{\mu\nu} + P^L_{\mu\nu} = P_{\mu\nu}\,,\qquad
P_{\mu\nu} = \delta_{\mu\nu} - \frac{p_\mu p_\nu}{p^2}\,,
\label{eq:projectors}
\eeqn
where $P_{\mu\nu}$ is the standard $O(4)$--symmetric
projector corresponding to the zero temperature case with
$$p^2 = \bp^2 + p_4^2\,.$$

At finite temperature the propagators $D_T$ and $D_L$ are, in general,
different from each other. The difference between spatially longitudinal and
spatially transverse functions arises due to interactions among the fields
while the free gauge theory has these functions equal:
$D_T^{\mathrm{free}} = D_L^{\mathrm{free}}$.
The interactions which lead to the difference between the propagators
$D_T$ and $D_L$ may be perturbative, as, for example,
in quantum electrodynamics~\cite{ref:Kapusta}, or these interactions
can be of purely nonperturbative origin, as, for example
in an Abelian gauge theory with contains only a compact gauge field~\cite{ref:cQED}.
The compact Abelian gauge theory possesses nonperturbative topological defects,
monopoles,
which affect drastically the propagator properties in the confining phase of the
theory.

In $SU(2)$ Yang--Mills theory the propagators $D_T$ and $D_L$ were
investigated using both numerical simulations on the
lattice~\cite{ref:structure:num:Cucch} and analytical calculations
in the continuum~\cite{ref:structure:an:Maas,
ref:structure:an:Gruter,ref:structure:an:Wambach,ref:structure:an:Alkofer}.
In the limit of vanishingly low temperatures the thermodynamics of
Yang-Mills theory imposes a certain constraint on the infrared
critical exponents which characterize the infrared behavior of the
correlators~\cite{ref:VIZ}.

From Eq.~\eq{eq:propagator:Tneq0} one can derive the electric--magnetic asymmetry of
the $A^2$ condensate in terms of the two propagators:
\beqn
\Delta_{A^2}(T) & = & \frac{N_c^2 - 1}{24\, \pi^3} T \int \dd^3 \bp \sum_n
\Bigl[\frac{3 \bp^2 - \omega_n^2}{\bp^2 + \omega_n^2} D_L(\bp,\omega_n)
\nonumber\\
& & - 2 D_T(\bp,\omega_n)\Bigr]
\label{eq:A2:SU2}
\eeqn
This expression should be ultraviolet finite up to a logarithmic renormalization.

\section{Numerical results in SU(2) Yang-Mills theory}
\label{sec:numerical_simulations}

In the following we report measurements of the electric--magnetic asymmetry
of the $A^2$ condensate that were performed in $SU(2)$ gluodynamics simulations.
The configurations have been created by means of a heatbath Monte Carlo code.
The minimal Landau gauge has been enforced for every
30-th configuration before measuring the $A_{\mu}^2$.
For the gauge fixing we have used the Simulated Annealing
algorithm~\cite{ref:bali-bornyakov}.
An interval of the gauge temperature ranging from $T_{\rm max}=1.0$ to
$T_{\rm min}=1.0 \cdot 10^{-5}$ has been traversed within 3000 sweeps with
linearly decreasing gauge temperature. This ``gauge cooling'' was followed
by obligatory overrelaxation until the required transversality was reached.
The stopping criterion was $\max_x \max_a |(\partial_\mu A^a_\mu)(x)| < 10^{-9}$.

We have evaluated the electric-magnetic asymmetry on lattices $16^3\times 4$,
$24^3\times 6$ and $32^3\times 8$. In the interval $\beta \in [2.20,2.95]$
we have selected a grid containing 51, 36 and 24 $\beta$-values, respectively.
In this way, an interval of physical temperatures $T \in [0.4 T_c,6.1 T_c]$
is covered. In the restricted range $T \in [0.4 T_c, 2.5 T_c]$, systems at
nearly equal temperatures are realized by lattices with different lattice
spacings. Close to the deconfining transition, the $3D$ volumes are equal
to each other with an aspect ratio of $N_{s}:N_{t}=4:1$.
The simultaneous evaluation of the asymmetry provides us with a
valuable assessment of potentially dangerous finite cutoff effects.
This does not seem to be a problem at all. As we will see later, data from
different lattice sizes are smooth and can be fitted simultaneously as
functions of the physical temperature.

The number of configurations at each combination of lattice size and $\beta$-value
was adjusted in such a way to give reasonable statistical error bars (typically,
the errors are of the order of a few percent near the phase transition temperature).
Close to the transition we needed about 1500 configurations per $\beta$ value
at the smallest lattice, and we found about 50 configuration per value of $\beta$
sufficient at the largest lattice far from the transition.

All our production measurements have been done with one Gribov copy only.
We stress that the Simulated Annealing algorithm already shifts the outcome
of a single gauge fixing closer to the (unknown) absolute maximum of the gauge
functional than several repetitions of pure overrelaxation could do.
We have actually checked the Gribov copy dependence at our middle-sized lattice,
$24^3 \times 6$ for a representative set of four temperatures corresponding
to the confinement region ($\beta=2.30$, $T \approx 0.65 \, T_c$),
to the deconfinement region ($\beta=2.7$, $T \approx 2.4 \, T_c$) and
to two temperatures close to the phase transition,
on the confinement ($\beta=2.40$, $T \approx 0.9 \, T_c$)
and on the deconfinement ($\beta=2.50$, $T \approx 1.25 \, T_c$)
sides. We took the first ten Gribov copies, $N_G=1 \dots 10$, by repeating the
Simulated Annealing gauge fixing. To form the ensemble collecting the best gauge
copies for each Monte Carlo configuration after $N_G$ repetitions of Simulated
Annealing we sampled the ``currently best'' copy among the preceding $N_G$
gauge-fixing trials. We found that -- within our statistical errors -- the
plateau value (for the ensembles with $N_G \to \infty$) of the electric-magnetic
asymmetry is already reached in the ensemble of gauge-fixed copies corresponding
to $N_G \simeq 4$ trials.
In the deconfinement phase the copy dependence is negligible since the
systematical uncertainty due to the Gribov copy dependence is much smaller
than the statistical errors of our calculations. In the confinement region
the uncertainty -- calculated as the relative deviation
of the measured value of the electric-magnetic asymmetry after $N_G$ trials
from the $N_G \to \infty$ plateau value --
is about $2\%$ while in the close neighborhood of the phase transition
the systematic correction to the electric-magnetic asymmetry due to Gribov
copy dependence may reach $10\%$. The asymmetry is slightly rising with
the number of gauge copies being under inspection. However, all characteristic
features of the asymmetry discussed in this article are unaffected by this
Gribov copy dependence.

The vector potential is extracted from the links as
\beqn
g A^a_{\mu}(x+\hat{\mu}/2) =
~{\mathrm{Tr}}\left[\tau_a~\left( U_{x,\mu} - U^{\dagger}_{x,\mu} \right)/(2~i~a) \right] \, .
\label{eq:Amu}
\eeqn

We express all dimensional quantities in units of the critical temperature.
According to Ref.~\cite{ref:critical:Tc:SU2}
the phase transition -- which is of the second order in
pure $SU(2)$ gauge theory -- happens at the critical temperature
\beqn
T_c = 0.694(18) \sqrt{\sigma(T=0)} = 305(8)\, \mbox{MeV}\,.
\eeqn

\begin{figure}[!htb]
\begin{center}
\includegraphics[width=8cm]{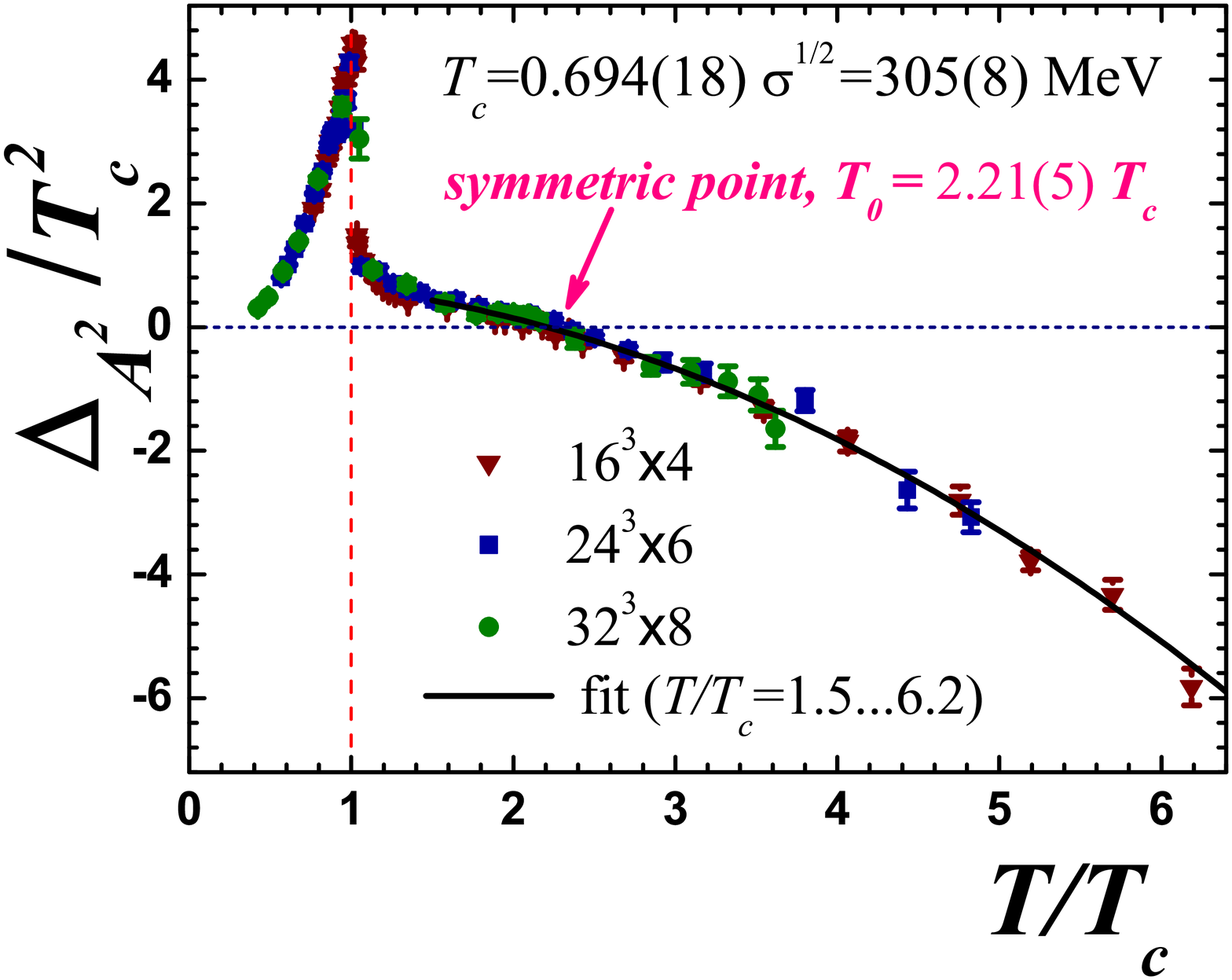}
\end{center}
\vskip -8mm
\caption{(Color online) The electric-magnetic asymmetry of the
$A^2$-condensate~\eq{eq:CA:definition} for $SU(2)$ gauge theory
normalized by the critical temperature squared
as function of $T/T_c$.
The high-temperature fit~\eq{eq:fit:highT} with the best fit
parameters~\eq{eq:best:fit:highT} is shown by the solid line.
The horizontal dashed line corresponds to $\Delta_{A^2} = 0$
while the vertical dashed line marks the critical temperature.
The symmetric point is explicitly indicated.}
\label{fig:CA:YM}
\end{figure}

The electric-magnetic asymmetry of the $A^2$-condensate~\eq{eq:CA:definition}
is shown in Figure~\ref{fig:CA:YM}.
One can observe a very good scaling: the results obtained
at various lattice sizes describe the same curve
when plotted in physical units. This fact indicates that
the lattice ultraviolet artifacts are negligible in
our numerical setup.

Before going into details we notice immediately
two distinct features of the temperature dependence of the
$A^2$ condensate asymmetry:
\begin{itemize}
\item[$\blacktriangleright$]
The maximum is taken at $T = T_{\mathrm{max}} \approx T_c$.
We observe a sharp maximum of $\Delta_{A^2}$ signalling that the asymmetry of the
gluonic medium is peaked around the phase transition.
\item[$\blacktriangleright$]
There is a symmetric point, $T = T_0 \approx 2 T_c$,
where the asymmetry vanishes. This point is realized in the deconfinement
phase sufficiently far from the phase transition
at a temperature approximately equal twice the critical temperature.
\end{itemize}
The mentioned points divide the phase diagram
into three separate regions:
\begin{itemize}
\item[$\blacktriangleright$]
Region 1: The confinement phase, $T \lesssim T_c$.
The asymmetry is a positive monotonically increasing
function of temperature in the confinement region.
\item[$\blacktriangleright$]
Region 2: The deconfinement phase at relatively low temperatures,
$T_c \lesssim T < T_0$. Here the asymmetry is still positive valued
and monotonically decreasing function of temperature.
The asymmetry vanishes at the temperature $T_0$, which we estimate below.
\item[$\blacktriangleright$]
Region 3: The deconfinement phase at high temperatures, $T> T_0$.
Here the asymmetry is negative valued and monotonically further decreasing
as a function of temperature.
\end{itemize}

As we speculate below in Section~\ref{sec:monopoles}, these regions are
characterized each by a particular dynamics of the Abelian monopoles.
These are singular configurations of the gluonic fields responsible for the
confinement of color in the low-temperature phase.
Region~1 corresponds to the phase, where the monopoles are condensed.
At higher temperatures, in Region~2, the monopole condensate melts into
a monopole liquid~\cite{Chernodub:2006gu}, whereas
at even higher temperatures, in Region~3, the liquid of the
Abelian monopoles is suggested to evaporate into a
gaseous state~\cite{Chernodub:2006gu}.
The transition between Region 1 and Region 2 is a true phase transition
(turning the monopole condensate into a monopole liquid), whereas
the transition between Region 2 and Region 3 (the evaporation of the monopole
liquid) is suggested to be a broad crossover. The interaction with electric
charges plays an important r\^ole in the dynamics of
monopoles~\cite{ref:Shuryak,ref:Shuryak:liquid},
such that we expect an effect of the changing monopole dynamics
on the electric-magnetic asymmetry of the condensate.

The deconfining transition at $T=T_c$ has a noticeable effect on the
asymmetry of the condensate since at this temperature the electric part of the
condensate is maximally dominating over the magnetic one.
This dominance is rapidly decaying just above $T_c$.
We define the maximum as
\beqn
\Delta^{\mathrm{max}}_{A^2} \equiv \max_T\,\Delta_{A^2}(T) = \Delta_{A^2}(T_{\mathrm{max}})\,.
\eeqn
We estimate the numerical value of the maximum as
\beqn
\Delta^{\mathrm{max}}_{A^2} & = & {[2.26(4) \, T_c]}^2 \equiv {[690(22)\,\mbox{MeV}]}^2 \,,
\label{eq:Delta:max}
\eeqn
and the temperature
\beqn
T_{\mathrm{max}} & = & \ 1.00(3) \, T_c \ \ \equiv \, 305(12)\, \mbox{MeV}\,.
\label{eq:T:max}
\eeqn

The symmetric point is realized at a temperature at which the
asymmetry of the condensate is vanishing. Our estimate of the symmetric point is
\beqn
T_0 = 2.21(5) T_c = 675(23)\,\mbox{MeV}\,,\qquad \Delta_{A^2}(T_0) = 0\,.
\label{eq:T0}
\eeqn
The change of sign of $\Delta_{A^2}$ happens in the deconfinement phase at
a temperature approximately twice the deconfinement temperature. In order
to accurately estimate the position of the symmetric point~\eq{eq:T0} we
performed a specific fit of the asymmetry throughout the deconfinement region.

The {\it high-temperature} fit is done with the help of the following fitting
function
\beqn
\Delta_{A^2}^{\mathrm{fit}} = \Delta_{A^2}^{(0)} - f \, T^2\,,\qquad \mbox{[at high temperatures]}
\label{eq:fit:highT}
\eeqn
where $\Delta_{A^2}^{(0)}$ and $f$ are two fitting parameters. The form of the fitting function
is inspired by the analytical examples provided by the photodynamics~\eq{eq:CA:photon}
and by the Abelian Higgs model~\eq{eq:AHM:lim:high} in their high temperature limits.
Surprisingly, the fit works very well not only at
very high temperatures, but it also works down to the temperatures as low as
$1.5\, T_c$. We obtain -- with a $\chi^2_{\mathrm{d.o.f.}} \approx 2$ --
the following best fit parameters:
\beqn
\Delta_{A^2}^{(0)} & = & {[0.894(14) \, T_c]}^2 = {[274(8) \, \mbox{MeV}]}^2\,,
\nonumber\\
f & = & 0.164(4)\,.
\label{eq:best:fit:highT}
\eeqn
The fit is shown in Figure~\ref{fig:CA:YM} by the solid line. According to the
high-temperature limits in
photodynamics~\eq{eq:CA:photon} and in the Abelian Higgs model~\eq{eq:AHM:lim:high},
one could have expected that each color component of the gluon would give the same
contribution $g^2/12$ to the coefficient $f$ in Eq.~\eq{eq:best:fit:highT}.
For $N_g = N^2_c - 1 = 3$ free gluons this coefficient should be equal to
$f=g^2/4$, indicating that
$g \lesssim 1$. This result is expected since in the considered
temperature regime the theory is still in a strongly
non-perturbative regime. One should note that the quadratic
fit~\eq{eq:fit:highT} may mimic some other, nontrivial $T$
dependence of the asymmetry, the exact form of which is difficult to
figure out at our accuracy. The fit is convenient for the estimation
of the symmetric point~\eq{eq:T0} which follows from
Eq.~\eq{eq:fit:highT} and Eq.~\eq{eq:best:fit:highT}.

\begin{figure}[!htb]
\begin{center}
\includegraphics[width=8cm]{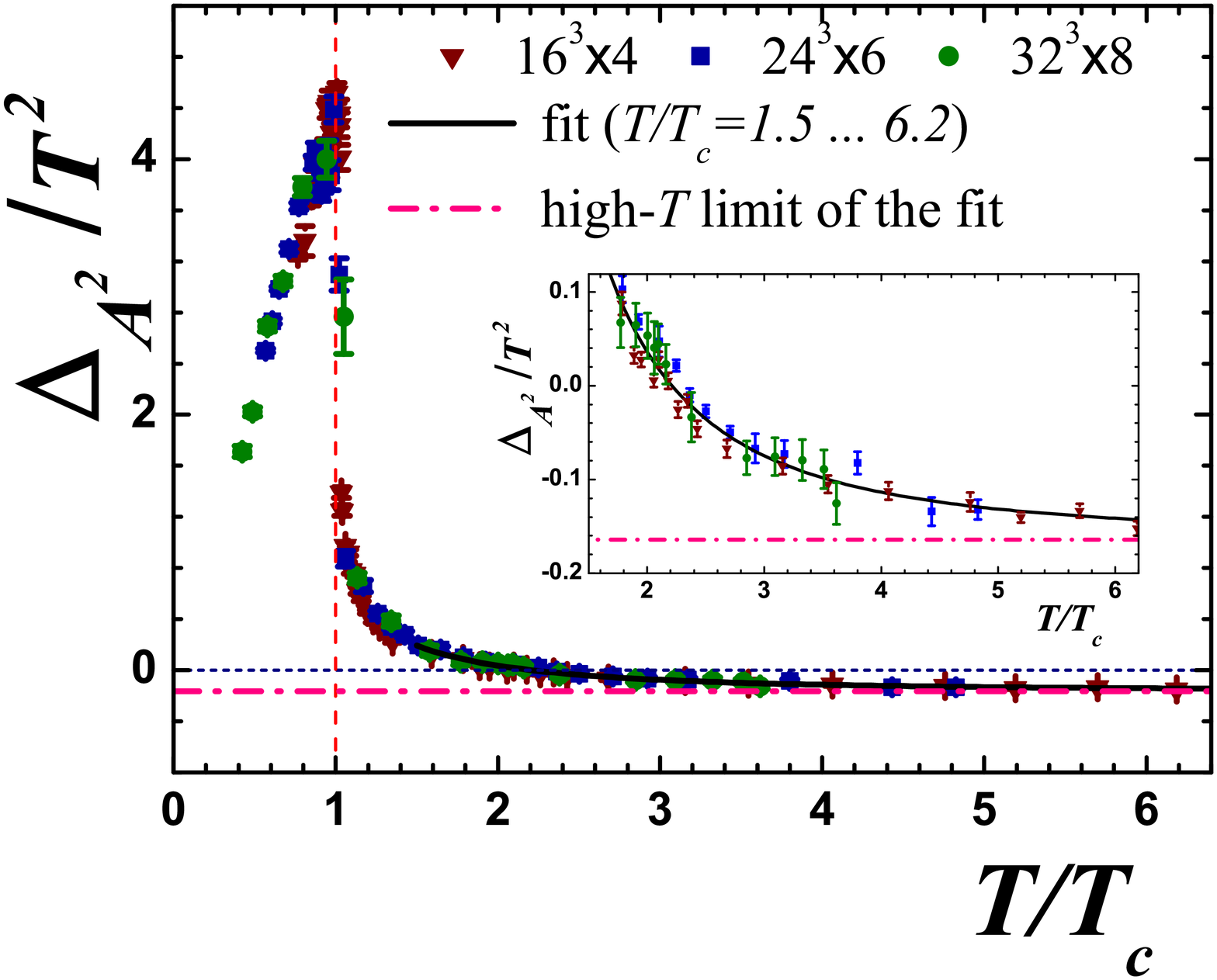}
\end{center}
\vskip -8mm
\caption{(Color online) The same as in Figure~\ref{fig:CA:YM} but for electric-magnetic
asymmetry of the $A^2$-condensate~\eq{eq:CA:definition}
normalized by the temperature $T^2$. The inset zooms in the high-temperature
region of the fit. The peaked value at $T=T_c$, Eq.~\eq{eq:best:fit:highT},
and the asymptotic value at $T \to \infty$, Eq.~\eq{eq:Delta:max}
are indicated by vertical dotted and horizontal
dash-dotted lines, respectively.}
\label{fig:YM:norm}
\end{figure}

In order to emphasize the approach of the electric-magnetic
asymmetry of the condensate to the asymptotic behavior at high
temperatures we present in Figure~\ref{fig:YM:norm} the asymmetry
normalized by the temperature squared.

The {\it low-temperature} limit
is especially interesting in view of
the existence of two different options: although Yang--Mills theory possesses
the mass gap, the asymmetry would not necessarily be
exponentially suppressed, as suggested in Eq.~\eq{eq:asymptotics}. Indeed,
the asymmetry could
behave polynomially, according to our calculation~\eq{eq:AHM:lim:low} in the broken
phase of the Abelian Higgs model. In order to figure out the behavior of the
asymmetry we have made an interpolating fit, which includes both options:
\beqn
\Delta_{A^2}^{\mathrm{fit}} = C_\Delta\, T_c^2 {\Bigl(\frac{T}{T_c}\Bigr)}^\nu \, \exp\{- m/T \}\,,\qquad \mbox{[at low $T$]}\,.
\label{eq:fit:lowT}
\eeqn
The polynomial behavior would be realized if $m=0$, while
the exponential suppression is in effect with $m \neq 0$.

\begin{figure*}[!htb]
\begin{center}
\begin{tabular}{cc}
\includegraphics[width=8cm]{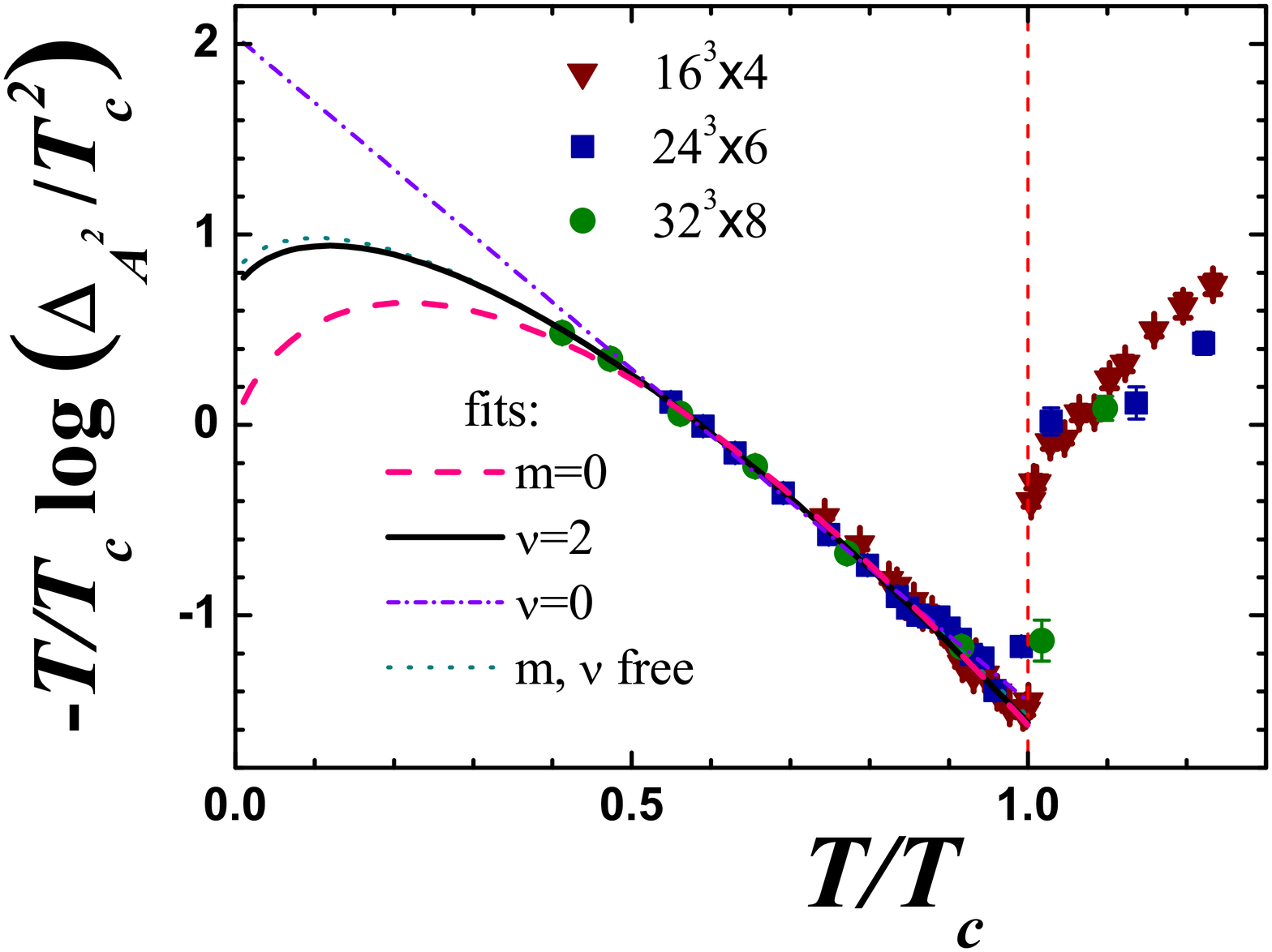} &
\includegraphics[width=8cm]{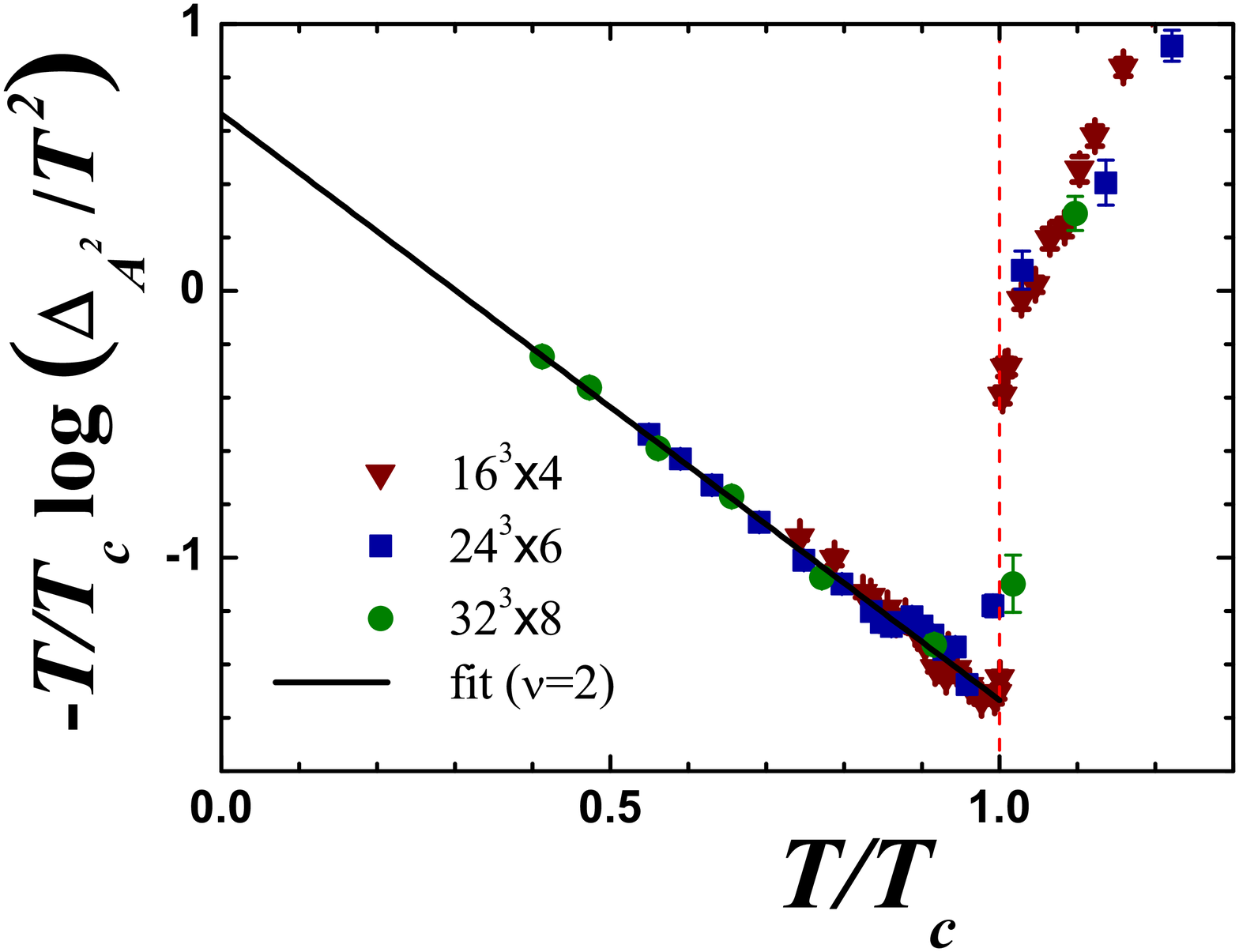} \\
(a) & (b)
\end{tabular}
\end{center}
\caption{(Color online) The behavior of the electric-magnetic asymmetry in the low-temperature
region. The lines represent the fits by the function~\eq{eq:fit:lowT} which
are discussed in the text.
Notice the different normalization of the expression under the logarithm:
in (a) the normalization factor is $1/T_c^2$ while in (b) this factor is $1/T^2$.}
\label{fig:fits:lowT}
\end{figure*}

We show the low-temperature asymmetry in Figures~\ref{fig:fits:lowT}(a) and (b).
In order to visualize the
exponential behavior in Figure~\ref{fig:fits:lowT}(a) we took the logarithm
of the asymmetry normalized
by the dimensional factor $T_c^2$ and then multiplied the logarithm by $- T/T_c$.
If the behavior of the asymmetry is proportional to the Boltzmann-like exponential
function (without the polynomial prefactor)
then the data must be linear in the low-temperature region. This is indeed
the case in the region close to the phase transition while as temperature gets
smaller then the deviation from the linear behavior becomes more visible.
Figure~\ref{fig:fits:lowT}(b) corresponds to the different normalization
factor under the logarithm, $T^2$.

First, we performed the fit using all three parameters $C_\Delta$, $\nu$ and $m$.
The best fit parameters are presented in Table~\ref{tbl:lowT}, and the
corresponding curve is shown in Figure~\ref{fig:fits:lowT}(a) by
the dotted line.
The fit was performed using all available values of the asymmetry in the
low-temperature region, $T < T_c$.
As one can see from the Table, the exponent $\nu$ is quite close to 2,
so that we have fixed $\nu=2$ and performed another fit. The quality of
both fits are characterized by almost the same value of $\chi^2/{\mathrm{d.o.f.}}$,
also presented in Table~\ref{tbl:lowT}. The fitting curve is shown in
Figure~\ref{fig:fits:lowT}(a) by the solid line.
Despite very good visual coincidence of the numerical data and the fitting curves, the high value of
$\chi^2/{\mathrm{d.o.f.}} \approx 6$ is due to the fact that the data corresponding to different lattice volume are
somewhat scattered with respect to each other in the region close to the phase
transition. Since the transition is of second order, we attribute the high
value of the $\chi^2/{\mathrm{d.o.f.}}$
parameter to finite-volume effects.

\begin{table}[htb]
\begin{tabular}{|l|l|l|l||l|l|}
\hline
$\nu$       & $m/T_c$       & $C_\Delta$ & $\chi^2/{\mathrm{d.o.f.}}$ & $m$, MeV & $1/m$, fm \\
\hline
1.82(28)    & 0.80(19)  & 10.24(1.96)   & 6.6 & 244(58) & 0.80(19)\\
2 [exact]   & 0.66(2)   & 9.00(32)    & 6.2 & 201(8)  & 0.98(4)\\
0 [exact]   & 2.04(3)   & 32.96(64)   & 14. & 622(18) & 0.32(1)\\
2.97(3)     & 0 [exact] & 4.84(32)    & 9.6 & 0 [exact]& $\infty$\\
\hline
\end{tabular}
\caption{The best fit parameters $\nu$, $m$ and $C_\Delta$ and the
parameter $\chi^2/{\mathrm{d.o.f.}}$
describing the electric-magnetic asymmetry by the fitting
function~\eq{eq:fit:lowT} in the low-temperature region.
The mentioned fitting curves are shown in Fig.~\ref{fig:fits:lowT}.
We also indicate the masses and the correlation
lengths $\lambda = 1/m$ in physical units.}
\label{tbl:lowT}
\end{table}

The polynomial prefactor plays an important r\^ole. If we set $\nu=0$,
the fitting function reduces to a purely exponential behavior.
In this case the quality of the fit deteriorates drastically.
The best fitting curve is shown in Figure~\ref{fig:fits:lowT}(a) by
the dotted curve. Similarly,
if we set the mass to zero, $m=0$, and fit the data using only $C_\Delta$
and $\nu$ as free parameters,
the quality of the fit gets worse, represented by the dashed line in
Figure~\ref{fig:fits:lowT}(a).
Moreover, according to Table~\ref{tbl:lowT} in this case the polynomial
behavior would be -- with a good accuracy --
proportional to $T^3$ and not to the fourth power
as one might have guessed from the example of the massive Abelian vector
model~\eq{eq:AHM:lim:low}. Thus, the most
plausible fit is characterized by an exponential suppression with a quadratic
prefactor:
\beqn
\Delta_{A^2}^{\mathrm{fit}} = C_\Delta\, T^2 \, \exp\{- m/T \}\,,
\label{eq:A2:2}
\eeqn
with
\beqn
C_\Delta =9.00(32)\,,\qquad m = 201(8)\, \mbox{MeV}\,.
\label{eq:A2:best:low}
\eeqn
The best fit curve corresponding to the quadratic polynomial
behavior~\eq{eq:A2:2} is also shown
in Figure~\ref{fig:fits:lowT}(b). We show the specific function of the data,
$-(T/T_c) \log \Delta_{A^2}/T^2$.
This function must be linear if the data obey the law~\eq{eq:A2:2},
and this seems to fit the data.
However, in order to figure out this fact with confidence
one needs to calculate the electric-magnetic asymmetry
at lower temperatures (the lowest temperature available to us
in our simulation was $T\approx 0.4 T_c$).

There are two important remarks now in order.
\begin{itemize}

\item[$\blacktriangleright$] Firstly, our data suggest that the behavior of
the electric-magnetic asymmetry at low temperature is not proportional to
the fourth power of temperature, as one could guess
from the simple case of a massive photon~\eq{eq:AHM:lim:low}. The behavior is
rather exponential~\eq{eq:A2:2} with a polynomial prefactor. According to our
discussion at the end of Section~\ref{sec:AHM} a purely polynomial behavior is
guaranteed if the propagator in momentum space tends to a certain non-vanishing
limit with vanishing momentum. The propagator in the Abelian Higgs model
provides us with a clear example of such a behavior.
According to our discussion, in the Abelian Higgs model the very reason
of the purely polynomial behavior is the appearance of the
``longitudinal'' massless pole,
$1/p^2$, in the integrand~\eq{eq:b2} of the integral representation of the
asymmetry~\eq{eq:inter:a2}. The $1/p^2$ term
enters this representation in the
combination $D(p^2)/p^2$, and if the propagator,
$D(p^2)$, would vanish at low momenta as, for example, $D(p^2) \propto p^2$,
then the polynomial behavior of the asymmetry would change to an exponential one.

Thus, the low-temperature behavior of the electric-magnetic asymmetry of the
dimension-2 condensate in Yang--Mills theory may shed some light on the
low-momentum behavior of the gluon propagator in the momentum space.

Indeed, the integral representation of the $A^2$-condensate in the Abelian Higgs model~\eq{eq:inter:a2} is similar
to the one in Yang-Mills theory~\eq{eq:A2:SU2},
such that the same considerations may apply.
The exponential suppression of the condensate at low
temperatures may signal that the thermal gluon propagator at low momenta
behaves softer than the tree-level propagator.
We should admit that we do not have data at low enough temperatures to prove
this fact firmly.

\item[$\blacktriangleright$] The second interesting observation is that the
mass which governs the exponential falloff of the asymmetry at low temperatures
is not the glueball mass, as na\"ively expected~\eq{eq:mass:naive}. According
to Eq.~\eq{eq:A2:best:low} this massive parameter is much smaller than the
glueball mass~\eq{eq:glueball},
namely of the order of $\Lambda_{\mathrm{QCD}}$, in other words,
of the order of the critical temperature, $T_c$, Eq.~\eq{eq:A2:best:low}.
Thus, the characteristic length
that describes non-perturbative effects related to the $A^2$-condensates
may be as large as one fermi. This could explain the fact that the exploration
of the low-momentum asymptotics of the gluon propagator
requires relatively large lattices~\cite{ref:huge:lattices}.

\end{itemize}

\section{Discussion: Confinement and Asymmetry of condensate}
\label{sec:monopoles}

It is extremely interesting to understand a physical reasons behind
the observed behavior of the electric-magnetic asymmetry,
Figure~\ref{fig:CA:YM}.
First of all it is striking that in the confining region the
$A^2$ condensate is {\it not} electric-magnetic symmetric.
Instead, the asymmetry is positive and is growing with temperature up
to its maximum which is realized just at the deconfinement phase transition.
We know from the considered example of the Abelian Higgs model that a
finite mass gap alone could not result in a
positive value of the asymmetry.
This suggests that the asymmetry of the dimension-2 condensate ought to
be related to the purely confining properties of the system.
This observation is in agreement with the original suggestion made in
Refs.~\cite{ref:Gubarev:Zakharov:PRL,ref:Gubarev:Zakharov:PLB}, as well
as with numerical results of Ref.~\cite{ref:Suzuki}, in which a relation
between the dimension-2 condensates and the confining string was
discussed. Thus, one can conclude that color confinement (and its agents)
may contribute to the unexpected behavior of asymmetry in the confinement phase.

It is generally accepted that the confinement of color can be explained
by the dynamics of either monopole-like gluonic configurations
(for a review see Ref.~\cite{ref:Review}) or by string-like
vortex configurations (a review can be found in Ref.~\cite{ref:review:vortices}).
In fact, these objects turn out to be interrelated
physically~\cite{ref:Griedt,ref:Zakharov,ref:boyko,ref:bornyakov}
and geometrically~\cite{ref:review:vortices}.
Confinement of color in the low-temperature phase is
caused by condensation of monopoles and by the spatial percolation of vortices.

Since the asymmetry of the $A^2$ condensate is presumably related to the color
confinement one should be able to trace the asymmetry back
to the dynamics of the monopoles (and of the center vortices, but below we
discuss the magnetic monopoles only). Moreover, the interactions of the electric
charges and the magnetic monopoles are important to determine the state
of the monopoles~\cite{ref:Shuryak,ref:Shuryak:liquid}.
Therefore, it is very natural to suggest that the monopole dynamics
leaves its foot-prints in the electric-magnetic asymmetry of the dimension-2
condensate.

We suggest that the increase of the electric-magnetic asymmetry in the
confinement phase is related to a gradual decrease of the monopole condensate,
as witnessed also by the decreasing string tension with $T \to T_c$ in $SU(2)$
gluodynamics.
Once the condensate has disappeared at $T = T_c$, the asymmetry starts to
become weaker.
According to Refs.~\cite{Chernodub:2006gu,ref:Shuryak,ref:Shuryak:liquid} in this
region the monopoles form a monopole liquid. The heating of the liquid leads
to its complete evaporation only at higher temperatures. According to
Ref.~\cite{Chernodub:2006gu} this should happen around $T \approx 2 T_c$
which is pretty close to the point $T_0$ where the asymmetry
vanishes~\eq{eq:T0}. Thus, one can suggest that at $T=T_0$ the liquid
predominantly turns into a monopole gas. The negative value of the
electric-magnetic asymmetry of the $A^2$ condensate is characterized by
the gaseous phase of the magnetic monopoles.

\section{Conclusion}
\label{sec:conclusions}

We have studied the electric-magnetic asymmetry of the $A^2$ condensate
in the Landau gauge in three gauge theories at finite temperature
in four space-time dimensions.

\begin{itemize}

\item[$\blacktriangleright$]
In photodynamics -- the free theory of a massless Abelian gauge
field~\eq{eq:photodynamics} -- the asymmetry can be computed
analytically. The asymmetry turns out to be negative for all
temperatures being proportional to the temperature
squared~\eq{eq:CA:photon}.

\item[$\blacktriangleright$]
In the Abelian Higgs theory~\eq{eq:AHM}, for the London limit
the analytical expression for the asymmetry is given by
Eqs.~\eq{eq:CA:AHM}, \eq{eq:energy} and \eq{eq:sigma}.
This theory corresponds basically to a free theory of
a massive gauge field. The asymmetry -- shown in Figure~\ref{fig:CA:AHM} --
is also negative for all temperatures, and the high- and low-temperature
limits can be computed, respectively, in Eqs.~\eq{eq:AHM:lim:high}
and \eq{eq:AHM:lim:low}.

\item[$\blacktriangleright$]
In the $SU(2)$ gauge theory we compute the electric-magnetic asymmetry
using numerical simulations on the lattice, Figure~\ref{fig:CA:YM}.
The temperature dependence of the asymmetry turns out to be unexpected:
at low temperatures the asymmetry turns out to be positive. It grows with
increasing temperature, reaching a maximum around the critical temperature
$T \approx T_c$. In the deconfinement phase the asymmetry drops rapidly with
increasing temperature.
At $ T= T_0 = 2.21(5) T_c = 675(23)\,\mbox{MeV}$
the asymmetry vanishes, and it becomes negative at higher temperature.

In a spirit of Ref.~\cite{ref:VIZ} we suggest that the low temperature
asymptotics of the $A^2$ condensate is related to the low-momentum behavior
of the gluon propagator. Our data at relatively low temperatures,
$0.4 T_c \lesssim T \lesssim T_c$, indicate that the asymmetry is suppressed
exponentially~\eq{eq:A2:2} providing strong arguments in favor of the
infrared suppression of the gluon propagator at low temperatures.

The mass, which governs the exponential low-temperature suppression of the
asymmetry~\eq{eq:A2:2}, is unexpectedly much smaller than the mass
of the glueball. We found $m = 201(8)\, \mbox{MeV}$, somewhat smaller than
the deconfinement temperature, $T_c$, of the pure Yang-Mills theory.
The corresponding correlations length, $\lambda = 1/m$ is of the order
of one fermi.
As a by-product, this result suggests that relatively large (many fermi in
one direction) lattice volumes are needed to
pin-down the low-momentum asymptotics of the gluon propagator.

\end{itemize}

The electric-magnetic asymmetry is most probably related to the changing
dynamics of the confining gluonic configurations, the magnetic monopoles.
In fact, the monopole dynamics may be foot-printed in the electric-magnetic
asymmetry of the condensate because the electric degrees of freedom affect
the properties of the magnetic monopoles~\cite{ref:Shuryak,ref:Shuryak:liquid}
and vice versa.
More in detail, the regions of
(1) positive and growing,
(2) positive and decreasing,
and (3) negative asymmetry coincide with the regions in which the monopoles
should, according to classification of Ref.~\cite{Chernodub:2006gu},
(1) be condensed, (2) form a liquid state and (3) form a predominantly
gaseous state, respectively.

\appendix

\section{The asymmetry in photodynamics}
\label{sec:appendix:photo}

The asymmetry of the condensate in the momentum representation is
\beqn
\Delta_{A^2}(T) & = &  T \sum_n \int \frac{\dd^3 p}{{(2\pi)}^3}
\Bigl[
\Bigl\langle A_4(\bp,\omega_n) A_4(-\bp,-\omega_n) \Bigr\rangle
\nonumber\\
& & \hskip -10mm - \frac{1}{3} \sum_{i=1}^3
\Bigl\langle A_i(\bp,\omega_n) A_i(- \bp, - \omega_n) \Bigr\rangle
\Bigr]
\label{eq:CA:sum}\\
&& \hskip -10mm \equiv
 T \sum_n \int \frac{\dd^3 p}{{(2\pi)}^3}
\Bigl[D_{44}(\bp,\omega_n) - \frac{1}{3} \sum_{i=1}^3 D_{ii}(\bp,\omega_n) \Bigr] \nonumber\\
&& \hskip -10mm \equiv
T \sum_n \int \frac{\dd^3 p}{{(2\pi)}^3} G_{A^2}(\bp,\omega_n) \, , \nonumber
\eeqn
where the momentum-dependent structure function of the asymmetry is
\beqn
G_{A^2}(\bp,\omega_n) = D_{44}(\bp,\omega_n)
- \frac{1}{3} \sum_{i=1}^3 D_{ii}(\bp,\omega_n) \, .
\eeqn

In order to proceed further we use the following trick.
It is well known~\cite{ref:Kapusta,ref:Kislinger} that sums over
Matsubara frequencies~\eq{eq:Matsubara} of the form~\eq{eq:CA:sum}
can generally be converted into integrals,
\beqn
& & \hskip -10mm T \sum_{n\in \Z} F(p_0 = i \omega_n) = \frac{1}{2 \pi i}
\hskip -2mm \int\limits^{+ i \infty + \epsilon}_{- i \infty + \epsilon} \hskip -3mm
\dd p_0 \, {\mathrm{Re}} \, F(p_0)
\nonumber\\
& & + \frac{1}{\pi i}
\hskip -2mm \int\limits^{+ i \infty + \epsilon}_{- i \infty + \epsilon} \hskip -3mm
\dd p_0 \, f_T(p_0) \, {\mathrm{Re}} \, F(p_0) \, ,
\quad p_0 = i p_4,
\label{eq:sum}
\eeqn
where $\epsilon \to + 0$, the real part of the function $F$ is defined as follows,
\beqn
{\mathrm{Re}} \, F(p_0) \equiv \frac{1}{2} \Bigl[F(p_0) + F(-p_0)\Bigr]\,,
\eeqn
and the temperature-dependent function
\beqn
f_T(p_0) = \frac{1}{e^{p_0/T} - 1}\,,
\label{eq:BE}
\eeqn
is the Bose-Einstein distribution of a bosonic particle
with the energy $\omega = p_0$ at temperature~$T$.
Equation~\eq{eq:sum} is valid for any analytical function $F(p_0)$ provided
it does not possess poles at imaginary $p_0$-axis.

It is easy to notice that the first term on the right hand side of Eq.~\eq{eq:sum}
is temperature-independent, and therefore it represents the zero-temperature part
of the sum. Since the summation over Matsubara frequencies is usually supplemented
with integration over the spatial momentum $\bp$, the zero-temperature part
can generally be divergent in the ultraviolet region. The second term is the
temperature-dependent correction which is usually finite because of the exponential
suppression of the ultraviolet modes with $p_0 \gg 0$.

The sum~\eq{eq:CA:sum} over the asymmetry structure function $G_{A^2}$,
\beqn
\Delta_{A^2}(T) = \Delta_{A^2}(T=0) + \delta \Delta_{A^2}(T) \; ,
\eeqn
with
\beqn
& & \hskip -5mm \Delta_{A^2}(T=0) =  \frac{1}{2 i} \int \frac{\dd^3 p}{(2\pi)^3}
\\
& &
\int\limits_{-i\infty+\epsilon}^{i\infty+\epsilon}
\frac{\dd p_0}{2\pi} \Bigl[C_A(\bp,i p_0) + C_A(\bp,-i p_0)\Bigr]
\nonumber
\eeqn
and
\beqn
& & \hskip -5mm \delta \Delta_{A^2}(T) = \frac{1}{2 i} \int \frac{\dd^3 p}{(2\pi)^3}
\\
& &
\int\limits_{-i\infty+\epsilon}^{i\infty+\epsilon} \frac{\dd p_0}{2\pi}
\Bigl[C_A(\bp,i p_0) + C_A(\bp, -i p_0)\Bigr] f_T(p_0) \; ,
\nonumber
\eeqn
is expressed through the simple function
\beqn
C_A(\bp,p_4) = \frac{\bp^2 - 3\, p_4^2}{3~(p^2)^2} \; .
\label{eq:integrandU1}
\eeqn

One can explicitly check that
\beqn
\Delta_{A^2}(T = 0) = 0 \,,
\eeqn
as we have expected. This result is intuitively clear
because at zero temperature the difference between
spatial and temporal directions disappears and therefore the space-time
asymmetry of any quantity must be zero. Thus, only the
temperature-dependent part of the $A^2$ condensates contributes to the
asymmetry.

At non-zero temperature we represent the asymmetry as the sum of the residues
\beqn
& & \Delta_{A^2}(T) = \delta \Delta_{A^2} (T)
\label{eq:inter:a2}\\
& & = \int \frac{\dd^3 p}{(2 \pi)^3} \sum_{p_0 > 0}
2~\res\,\Bigl({\mathrm{Re}}~C_A (\bp, i p_0)~f(p_0)\Bigr)\,.
\nonumber
\eeqn

The residue in the case of an $n$th order pole at $z = a$ is defined as
\beqn
\res\, f(a) = \frac{1}{(n-1)!} \lim_{z \to a} \frac{\dd^{n-1}}{\dd z^{n-1}}
\Bigl[(z-a)^n f(z)\Bigr] \; ,
\eeqn
such that
\beqn
{\mathrm {for}~} n=1 \qquad \res\, f(z) & = & \lim_{z \to a} \Bigl[(z-a) f(z)\Bigr]
\label{eq:res1}\\
{\mathrm {for}~} n=2 \qquad \res\, f(z) & = & \lim_{z \to a} \frac{\dd}{\dd z} \Bigl[(z-a)^2 f(z)\Bigr] \, .
\qquad \label{eq:res2}
\eeqn
The quantity in the integrand,
\beqn
2~\Bigl({\mathrm{Re}}~C_A (\bp, i p_0)~f(p_0)\Bigr) =
\frac{2}{3} \frac{(\bp^2 + 3 p_0^2)~f_T(p_0)}
                 {(p_0-|\bp|)^2 (p_0+|\bp|)^2} \, ,
\nonumber
\eeqn
has a double pole in the complex $p_0$ plane on the positive real axis at $p_0 = |\bp|$.
Using Eq.~\eq{eq:res2} and then Eq.~\eq{eq:inter:a2} we get Eq.~\eq{eq:CA:photon}
for the case of the free massless photon.

\break

\section{The asymmetry in the Abelian Higgs model}
\label{sec:appendix:AHM}

In the Abelian Higgs model the instead of Eq.~(\ref{eq:integrandU1}) we get
\beqn
C_A(\bp,p_4) = \frac{\bp^2 - 3\, p_4^2}{3~(p^2)~(p^2 + m^2)} \; .
\label{eq:integrandAHM}
\eeqn
and, therefore,
\beqn
& & 2~\Bigl({\mathrm{Re}}~C_A(\bp,i p_0)~f(p_0)\Bigr)
\label{eq:b2}\\
& & \hskip 20mm =
\frac{2}{3} \frac{(\bp^2 + 3 p_0^2) \, f_T(p_0)}
{[p_0^2 - \bp^2]~[p_0^2-(\bp^2+m^2)^2]} \, .
\nonumber
\eeqn
Thus there are two single ($n=1$) poles in the complex $p_0$ plane
on the positive real axis at $p_0 = |\bp|$ and $p_0 = \sqrt{\bp^2 + m^2}$.  Using Eq.~\eq{eq:res1}
we obtain
\beqn
& & \hskip -5mm \res \Bigl[2~\Bigl({\mathrm{Re}}~C_A (\bp,i p_0)~f(p_0)\Bigr) \Bigr]
\nonumber\\
& = &
\frac{4}{3 m^2} \Bigl[\sqrt{\bp^2 + m^2}~f_T(\sqrt{\bp^2 + m^2})
- |\bp|~f_T(|\bp|)\Bigr] \nonumber\\
& & - \frac{1}{3} \frac{f_T(\sqrt{\bp^2 + m^2})}{\sqrt{\bp^2 + m^2}} \, .
\eeqn
Substituting this result in Eq.~\eq{eq:inter:a2} we get Eqs.~\eq{eq:CA:AHM}, \eq{eq:energy} and \eq{eq:sigma}.

\begin{acknowledgments}

MNCh. was partly supported by grants RFBR 06-02-04010-NNIO-a,
RFBR 08-02-00661-a, a joint German-Russian grant DFG-RFBR 436 RUS,
a grant for scientific schools NSh-679.2008.2 and by the Federal
Program of the Russian Ministry of Industry, Science and Technology
No 40.052.1.1.1112 and by the Russian Federal Agency for Nuclear Power.
He appreciates a STINT Institutional grant IG2004-2 025.
EMI is supported by DFG through the Forschergruppe FOR 465.
He thanks his co-authors within the Adelaide-Berlin-Dubna-Moscow
``Infrared QCD'' collaboration, in particular Vitaly Bornyakov,
Valya Mitrjushkin, Michael M\"uller-Preussker, Lorenz von Smekal and
Andre Sternbeck, for cooperation and discussions on related questions.
\end{acknowledgments}


\begin{thebibliography}{99}
\bibitem{footnote1}
In this paper are working in $D=4$ dimensional Euclidean space-time
suitable for numerical lattice simulations of the $(3+1)$ dimensional $SU(2)$
gauge theory. The phase transition in this theory is of second order while
in the more interesting case of SU(3) gauge theory the phase transition if of a (weak) first order.
\end{thebibliography}
\end{document}